\documentclass{v23windows}

\def\be{\begin{equation}}
\def\ee{\end{equation}}
\def\bea{\begin{eqnarray}}
\def\eea{\end{eqnarray}}

\usepackage{float}
\usepackage{slashed}
\usepackage{caption}

\begin{document}
\vspace*{4cm}
\title{CELESTIAL OBJECTS AS DARK MATTER COLLIDERS}

\author{ THONG T. Q. NGUYEN}

\address{Institute of Physics, Academia Sinica,\\
 Nangang, Taipei 11529, Taiwan}

\maketitle\abstracts{
In the vicinity of the Milky Way Galactic Center, celestial bodies, including neutron stars, reside within a dense dark matter environment. This study explores the accumulation of dark matter by neutron stars through dark matter-nucleon interactions, leading to increased internal dark matter density. Consequently, dark matter annihilation produces long-lived mediators that escape and decay into neutrinos. Leveraging experimental limits from IceCube, ANTARES, and future projections from ARIA, we establish constraints on the dark matter-nucleon cross section within a simplified dark $U(1)_{X}$ mediator model. This approach, applicable to various celestial objects and dark matter models, offers insights into the intricate interplay between dark matter and neutron stars near the Galactic Center.}

\section{Introduction}
\label{sec:intro}
The nature of dark matter (DM) stands as a central enigma in particle physics, necessitating extensions beyond the Standard Model (SM)~\cite{Bertone:2004pz}. To elucidate its properties, physicists have undertaken a comprehensive experimental quest to explore DM's interaction with SM particles~\cite{Bertone:2018krk}. One intriguing hypothesis garnering recent attention posits that DM interacts with SM fields through a dark force carrier, potentially via direct interaction with the SM or through mixing with SM gauge or Higgs bosons. In scenarios with modest couplings to the SM, these particles can exhibit remarkable longevity, prompting the use of techniques designed for the detection of exotic, long-lived particles~\cite{FASER:2018eoc}. For even smaller couplings, the new force carriers might have lifetimes extending beyond planetary scales, resulting in relatively weaker constraints~\cite{Feng:2015hja}. In this parameter space, distant compact celestial objects emerge as potential DM sources, where they can accumulate their own DM clouds~\cite{Bertone:2007ae} and offering opportunities to search for annihilation signals through Indirect Detection methods.

Recently, research exploring this idea has expanded to various types of observable compact objects, including those visible to the naked eye, such as the Sun~\cite{Leane:2017vag} and Jupiter~\cite{Leane:2021tjj}, as well as celestial bodies near the Galactic Center, such as neutron stars, brown dwarfs~\cite{Leane:2021ihh}, and white dwarfs~\cite{Acevedo:2023xnu}, guided by gamma-ray experimental limits~\cite{Malyshev:2015hqa}. In this study, we utilize neutrino experimental limits from IceCube and ANTARES~\cite{ANTARES:2015moa}, along with the population data of neutron stars in the Galactic Center. Through this framework and employing a simplified dark vector mediator model, we establish constraints on the dark matter-nucleon cross section~\cite{Nguyen:2022zwb}. While some sensitivities have been previously estimated in this scenario~\cite{Bose:2021yhz}, our emphasis is on the long-lived parameter regime. Additionally, we consider the potential of future high-energy neutrino observatories, such as ARIA~\cite{Anker:2019rzo}, to extend this sensitivity in the future.

\section{Dark Matter capture and annihilation inside neutron stars}
\label{sec:DMcapt}
Celestial bodies, including neutron stars situated in dark matter-rich regions, present an especially enticing target for observation. These objects have the inherent ability to traverse through a denser dark matter environment compared to that in the solar system. This phenomenon occurs when dark matter interacts strongly enough with these celestial bodies, causing it to lose kinetic energy and become gravitationally bound. Consequently, these bodies accumulate a localized cloud of dark matter, significantly denser than the surrounding ambient density. Hence, the capture rate, indicating the extent to which dark matter can accumulate within stars, is contingent upon the properties of the celestial object, such as its mass, radius, and elemental composition, as well as the cross section between dark matter and the constituent components.

As the dark matter cloud accumulates around a celestial body, its constituents initiate annihilation processes. If these annihilations lead to standard model particles or mediators with short lifetimes, they introduce a new heating mechanism that could influence the energy balance of the celestial object. However, if the annihilation or decay results in dark mediator particles with sufficiently long lifespans to allow them to escape before decaying, yet decay before reaching Earth, it presents a unique opportunity to investigate weakly-coupled dark mediators. Neutron stars, expected to be abundant in highly dark matter-rich environments, could collectively generate a prominent dark matter annihilation signal, offering a potential avenue for exploration.

\begin{figure}[H]
\centering
\captionsetup{justification=centering}
\includegraphics[width=0.8\textwidth]{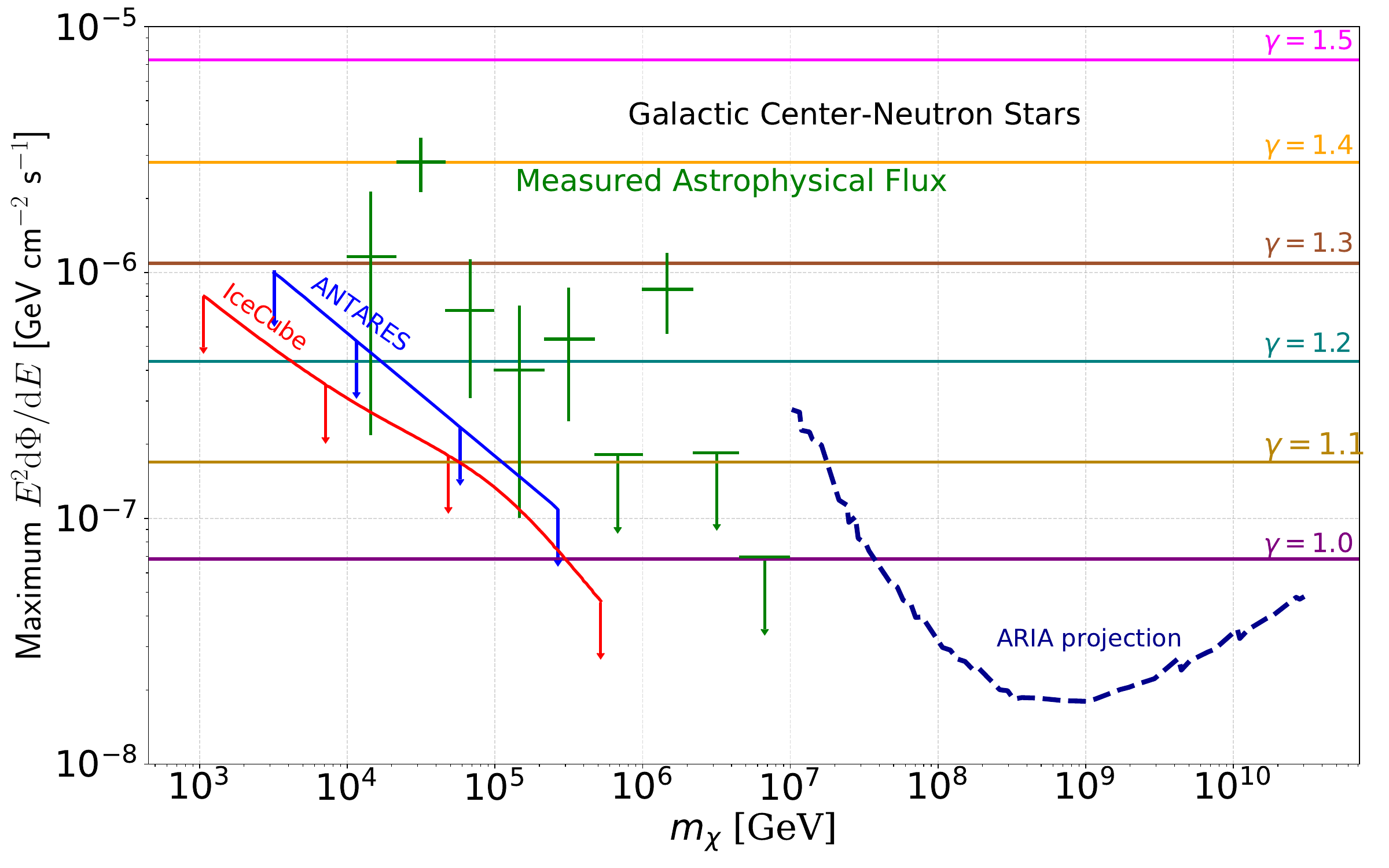}
\caption{Maximally optimistic values of differential flux
of muon neutrinos from Galactic Center neutron stars.
The IceCube 7-year upper limit is shown in red, the
ANTARES limit in blue, and the measured IceCube
diffuse flux in green. Future projected upper limits from
ARIA are indicated in dark blue.}
\label{fig:MaxFlux}
\end{figure}

We consider a specific simplified model for the Dirac
fermion dark matter $\chi$ which is a singlet under the SM
gauge interactions, but interacting with a mediator corresponding to a $U(1)_{X}$ gauge symmetry~\cite{Carena:2004xs}. The dark gauge boson $X_{\mu}$ picks up interactions with the SM via kinetic mixing:
\begin{equation}
\mathcal{L}\supset -\frac{1}{4}X_{\mu\nu}X^{\mu\nu}-\frac{\epsilon}{2}X_{\mu\nu}B^{\mu\nu}-\frac{1}{2}m_{X}^{2}X_{\mu}X^{\mu}+\bar{\chi}(i\slashed{D}_{U(1)_{X}}-m_{\chi})\chi,
\end{equation}
where $\epsilon$ parametrizes the kinetic mixing between $X_{\mu}$ and SM hypercharge interaction. We explore DM above the unitarity limit, typically necessitating a non-thermal production mechanism. Additionally, we investigate scenarios where dark matter annihilates into a pair of mediators with weak interactions with ordinary matter, enabling them to escape the celestial body's effective volume. Subsequently, these mediators decay into high-energy neutrinos. In Figure \ref{fig:MaxFlux}, we illustrate the maximum achievable signal for various selections of the dark matter density profile. Additionally, we incorporate the upper limits from IceCube, ANTARES, and the future projections from ARIA for comparison in this scenario.

\section{Bounds on dark matter-nucleon cross section}
Transitioning from the maximally optimistic scenario, we establish constraints on the cross section for spin-independent scattering of dark matter with neutrons. These constraints are delineated as a function of the dark matter mass, under the condition where $m_{X}\ll m_{\chi}$ and considering an appropriate mediator lifetime enabling escape from neutron stars and subsequent decay between the Galactic Center and our detectors. We adopt a conservative approach, assuming the entire neutrino flux originates from the dark matter signal while disregarding all other potential background sources.

\begin{figure}[H]
\centering
\captionsetup{justification=centering}
\includegraphics[width=\textwidth]{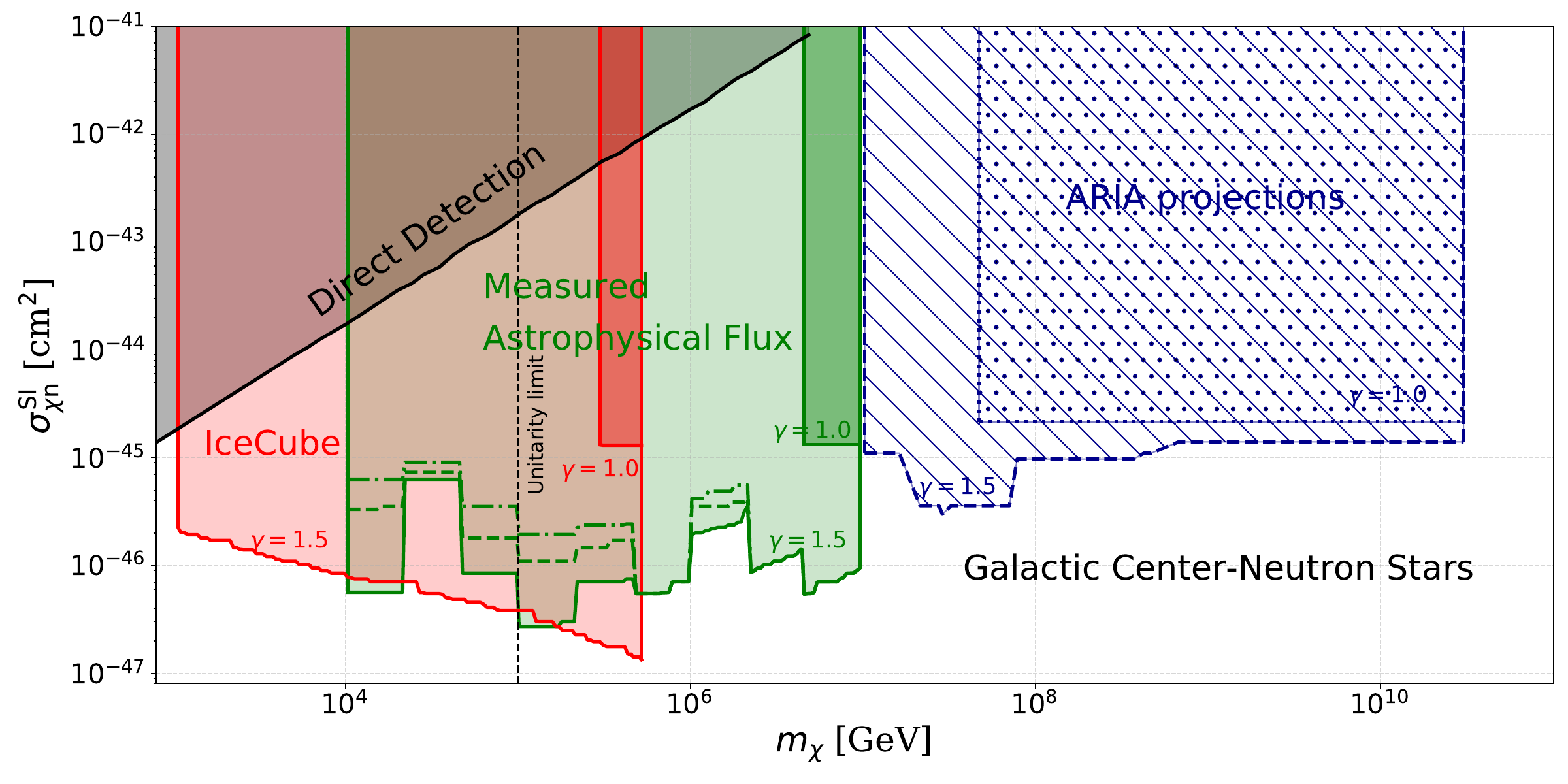}
\caption{Upper limits on the spin-independent cross section of dark matter scattering with neutrons, in the scenario
in which accumulated dark matter annihilates into mediators which escape the neutron star and subsequently decay
into neutrinos, for dark matter the generalized NFW profiles described by $\gamma=1.0 $ and $\gamma = 1.5$.}
\label{fig:xsect}
\end{figure}

In Figure \ref{fig:xsect}, we present the derived limits and projected constraints (for various dark matter profiles), juxtaposed with current limits from terrestrial dark matter searches for comparison~\cite{PICO:2019vsc}. These findings highlight the distinctive insights offered by existing high-energy neutrino observatories into theories involving dark matter with light, long-lived mediators decaying into neutrinos. Moreover, our results underscore the potential of future initiatives like ARIA, which have the capacity to significantly broaden our understanding and explore new realms within the dark matter parameter space.

\section{Summary}
In this paper, we have explored the feasibility of utilizing neutrinos generated from the decay of long-lived mediators, originating from the annihilation of dark matter clouds anticipated to amass around neutron stars in the Galaxy. The complexity of these theories poses significant challenges for terrestrial searches. Our findings, as summarized in Figure \ref{fig:xsect}, underscore the crucial role played by current high-energy neutrino observatories. Moreover, our study highlights the promising potential of future projects in shedding light on these intricate phenomena\cite{TTT}.

\section*{Acknowledgments}
This work is supported in part by the Ministry of Science and Technology (MOST) of Taiwan under Grant No 111-2112-M-001-035 and by the Sciences and Engineering Faculty of Sorbonne Université. T.T.Q.N. would like to thank the organizers of the conference for the opportunity to present our work, and for providing a venue for stimulating discussions with fellow participants.

\end{document}